\documentclass[twocolumn,preprintnumbers,superscriptaddress,amsmath,amssymb,prb]{revtex4-2}
\usepackage{graphicx}
\usepackage{dcolumn}
\usepackage{bm}
\usepackage{lipsum}  
\usepackage{hyperref}
\usepackage[mathlines]{lineno}

\usepackage{subcaption}
\usepackage{float} 
\usepackage{mathrsfs}
\usepackage{amsfonts}

\usepackage{xcolor} 
\usepackage{soul}   
\usepackage[utf8]{inputenc}  
\usepackage[T1]{fontenc}
\usepackage{amsmath}
\usepackage{units}

\usepackage{gensymb}
\usepackage{caption} 
\usepackage{newtxtext} 
\usepackage{newtxmath} 

\usepackage{bm}

\renewcommand{\mathbf}[1]{\bm{#1}}

\newcommand{\sout}[1]{}

\definecolor{lightblue}{rgb}{0.68, 0.85, 0.9}
\definecolor{lightgreen}{rgb}{0.56, 0.93, 0.56}
\UseRawInputEncoding

\newcommand{\zw}[1]{\textcolor{black}{#1}}
\newcommand{\rj}[1]{\textcolor{black}{#1}}
\newcommand{\rjb}[1]{\textcolor{black}{#1}}
\newcommand{\rjc}[1]{\textcolor{black}{#1}}

\begin{document}
\preprint{APS/123-QED}


\title{Stability and dynamics of magnetic skyrmions in  FM/AFM heterostructures}

\author{Rajgowrav Cheenikundil}
 \affiliation{School of Science and Technology, \"Orebro University,  SE-701 82, \"Orebro, Sweden }
 
\author{Zhiwei Lu}
 \affiliation{Department of Applied Physics, School of Engineering Sciences, KTH Royal Institute of Technology, SE-10691, Stockholm, Sweden}

\author{Manuel Pereiro}
\affiliation{Department of Physics and Astronomy, Uppsala University, SE-75120, Uppsala}%

\author{Anna Delin}
\affiliation{Department of Applied Physics, School of Engineering Sciences, KTH Royal Institute of Technology, SE-10691, Stockholm, Sweden}
\affiliation{\rjb{SeRC (Swedish e-Science Research Center), KTH Royal Institute of Technology}} 
\affiliation{\rjb{Wallenberg Initiative Materials Science for Sustainability (WISE), KTH Royal Institute of Technology}}

\author{ Danny Thonig}
\affiliation{School of Science and Technology, \"Orebro University, SE-701 82, \"Orebro, Sweden } 
\affiliation{Department of Physics and Astronomy, Uppsala University, SE-75120, Uppsala}%

\date{\today}

\begin{abstract}
Magnetic skyrmions have garnered attention for their potential roles in spintronic applications, such as information carriers in computation, data storage, and nano-oscillators due to their small size, topological stability, and the requirement of small electric currents to manipulate them. Two key challenges in harnessing skyrmions are the stabilization requirement through a strong out-of-plane field, and the skyrmion Hall effect (SkHE). Here, we present a systematic model study of skyrmions in FM/AFM multi-layer structures by employing both atomistic Monte Carlo and atomistic spin dynamics simulations. \rjb{We demonstrate that skyrmions stabilized by exchange bias have superior stability than field-stabilized skyrmions due to the formation of a magnetic imprint within the AFM layer. Additionally, stacking two skyrmion hosting FM layers between two antiferromagnetic (AFM) layers suppresses the SkHE and enables the transport of AFM-coupled skyrmions with high velocity in the order of a few Km/s}. This proposed multi-layer configuration could serve as a pathway to overcome existing limitations in the development of skyrmion-based devices, and the insights obtained through this study contribute significantly to the broader understanding of topological spin textures in magnetic materials.
\end{abstract}

\maketitle

\section{\label{sec:level1} Introduction}
Magnonics has emerged as a central area for the development of the next generation of low-power, high-speed computational devices. This progress is notably illustrated by the use of magnetic principles in magnetic RAMs (MRAMs), already in commercial use \cite{samsung2022}, showcasing the effectiveness of magnetic technologies in enhancing speed and reducing power consumption. In addition to this, the rapid advancement in artificial intelligence has intensified the search for low-powered artificial neural networks. \zw{Spintronics}-based neural networks have already been theoretically proposed \cite{Papp2021, Ross2023}, making the research into  \zw{spintronics} devices more relevant and timely \cite{EUProject793346, CORDIS793346}, especially considering the growing demand for energy-efficient AI computations. The exploration of \zw{spintronics} applications in neural networks could lead to significant breakthroughs in the field of AI, offering a promising avenue for the development of next-generation computing technologies.

{Magnetic skyrmion-based devices are proposed to be at the forefront of this technological shift} \cite{Muhl2009, Li2023}. Skyrmions have attracted significant attention as potential information carriers in \zw{spintronics} systems owing to their compact size \cite{Koshibae2015}, ability to be closely packed \cite{Sampaio2013, Zhang2015-edge-repulsion},  and exceptional stability. Their interaction with spin currents via spin transfer torque  \cite{Sampaio2013} offers an opportunity for precise control, a critical factor for practical applications in advanced computational systems. Moreover, skyrmions require a significantly lower critical electric current density for movement, usually five orders of magnitude smaller than the requirement for \rj{magnetic} domain walls \cite{Jonietz2010, Yu2012, Yu2020}. This reduced need for driving electric current is beneficial, as it minimizes energy loss due to dissipation and lowers the power requirements for these devices to two orders of magnitude less than what is needed for conventional CMOS devices. Skyrmions can also be controlled by other driving forces such as surface acoustic waves (SAW) \cite{Miyazaki2023}. Furthermore, skyrmions could be easily detected utilizing magneto-transport effects such as tunnel magneto-resistance and the non-collinear magneto-resistance arising due to band structure changes \cite{Crum2015, Hanneken2015}. They can also be identified with purely electrical techniques, making use of the topological Hall effect \cite{Hamamoto2016}. The ease of transport and measurement of skyrmions makes them ideal candidates for logic and computation applications. Designs and proposals for skyrmion-based devices are already emerging, including racetrack devices \cite{Tomasello2014, Yang2022, Morshed2022, He2023, Belrhazi2022}, logic gates \cite{Zhang2015}, and synapses \cite{chen2021surface}.

Our research explores skyrmion-based race-track type devices,  in which the skyrmions are moved in a controlled way with spin currents. The information bits are carried by a sequence of skyrmions moving along the track. There exist two significant challenges in the design of such a device.  The first challenge is the skyrmion Hall effect (SkHE) \cite{Jiang2016,juge2019}, where skyrmions when propelled by spin currents, exhibit a transverse motion instead of moving parallel to the current flow.  This transverse deviation results in the annihilation of the skyrmions at the edge of the track and hence it poses a significant difficulty. This effect arises as a counter-effect to the topological Hall effect experienced by the charge-carrying electrons, which in turn originates from the topology of the skyrmion. Various strategies have been explored, to address the skyrmion Hall effect, including geometric  \cite{Shigenaga2023, Pathak2021} and acoustic\cite{Chen_2023} control, and the use of skyrmionium structures \cite{Kolesnikov2018}. A promising approach is the coupling of two skyrmions with opposite Hall effects, which can lead to the cancellation of the SkHE \cite{Zhang2016, Silva2020, Jia2020, Xia2021_afm, Shen2019}, paving the way for more controlled skyrmion-based devices. The second challenge in the design of skyrmion racetrack devices is the requirement of a strong out-of-plane magnetic field to keep the skyrmions stable, making the design of such a device prohibitively complex. Various techniques have been employed to eliminate the requirement of an external magnetic field to keep the skyrmions stable. One of the promising approaches is to couple the skyrmion hosting ferromagnetic layer to an exchange-biased material, where the exchange bias will substitute the external magnetic field \cite{Yu2018, Rana2020}. Such multilayer setups were already experimentally realized and skyrmions were stabilized at zero fields in an ultra-thin FM/AFM heterostructure \cite{He2023_exb}. This achievement of experimentally realizing these setups marks a significant advancement. However, to broaden these results to more universal applications and to gain a deeper understanding of the intricate physical mechanisms involved, further research is necessary. Particularly, numerical simulation studies are essential to explore the interplay of various material properties on skyrmion stability and the strength of exchange bias in such systems. 
In this study, we focus on the stabilization of magnetic skyrmions by exchange-bias in ferromagnetic/antiferromagnetic (FM/AFM) multi-layer structures by employing atomistic Monte-Carlo and atomistic spin-dynamics simulations \cite{UppsalaUPASD}. {Our objectives include developing a model FM/AFM multilayer structure to study the dependence of exchange bias effect on various material and geometrical parameters. Additionally, we aim to investigate the stability of skyrmions within these multilayers under ambient conditions and explore the potential of using exchange bias to mitigate the skyrmion Hall effect, ultimately aiming to develop efficient skyrmion transport devices}.  

This paper is organized as follows. In Section II, the theory and techniques used for the numerical simulations are discussed in detail. In Section III, the results are presented in deductive order. Finally, in the last section, Section IV, we provide a summary and an outlook.

\section{\label{sec:level2} Methodology}

\subsection{Theory and modeling}
 The simulations are performed using the atomistic simulation tool UppASD \cite{UppsalaUPASD, Eriksson2016-ma}. Utilizing an atomistic model allows an accurate simulation of the magnetic system \rj{when compared to continuum models}, as this approach enables detailed calculation of the intricate interactions among atoms and considers factors like interracial effects,  \rj{antiferromagnetic coupling between neighboring atoms},  and localized properties, such as anisotropies.
 
 In this study, the properties of the system are modeled by the Hamiltonian given by

\begin{equation}
 \begin{aligned}
\mathscr{H}  = & - \frac{1}{2}\sum_{i \neq j} J_{ij} \mathbf{S}_i \cdot \mathbf{S}_j - \frac{1}{2}\sum_{i \neq j} \mathbf{D}_{ij} \cdot (\mathbf{S}_i \times \mathbf{S}_j) \\
  & -  \sum_i  K_{i}  ( \mathbf{S}_{i}  \cdot  \mathbf {e}^k_i  )^2 - \mu_{0} \sum_i \mathbf {H}_{\text{ext}} \cdot \mathbf {S}_{i}
 \end{aligned}
 \label{hamiltonian}
\end{equation}
where $\mathbf{ S}_i $ is the spin vector, ${ J_{ij}}$ is the isotropic part of the exchange coupling between the $i^{th}$ and $j^{th}$ atomic spin, $J > 0$ will result in ferromagnetic coupling and $J < 0$ will give antiferromagnetic coupling. \rjc{The breaking of the inversion symmetry at the FM/AFM interface could also lead to the rise of anisotropic terms of the symmetric exchange. We do not consider the contribution of these interactions in our Heisenberg term.} $\mathbf{D}_{ij}$ is the Dzyaloshinski Moriya interaction (DMI) vector between the $i^{th}$ and $j^{th}$ spins. In our model, we assumed  $\mathbf{D}_{ij}$ to have a direction parallel to the distance vector \zw{$\mathbf{r}_{ij}= \mathbf{r}_j - \mathbf{r}_i$} between the position $\mathbf{r}_i$ and $\mathbf{r}_j$ of the $i^{th}$ and $j^{th}$atom, respectively, which is a scenario commonly observed in bulk materials with non-centrosymmetric crystal structure \cite{Muhl2009, Nagaosa2013} \rjc{when there is an n-fold symmetry along the bond direction \cite{moriya1960}}.

\rj{$K_{i}$ is the uniaxial anisotropy constant and  $\mathbf {e}^{k}_{i}$ is the direction of easy axis for the  $i^{th}$ atom. The constant $\mu_0$ is the permeability of free space}. The external magnetic field is given by $ {\mathbf H}_{\text{ext}}$. The temperature-induced effects of the system can be simulated using Monte Carlo simulations. \rjb{Here, the parametrized energy given by Eq.\eqref{hamiltonian} was then used to obtain the thermal equilibrium state as a function of temperature and external magnetic field.} \cite{Eriksson2016-ma} \sout{and it is also used to obtain the equilibrium ground state of the system under a given set of conditions, \rj{such as temperature, field strength, and initial magnetic state}}. The dynamics of the spins are governed by the Landau-Lifshitz-Gilbert equation (LLG), given by \cite{Carpio2016MagnetizationDO},

\begin{equation}
\begin{aligned}
\frac{\partial \mathbf{S}_i}{\partial t} = & -\gamma \mathbf{S}_i \times \mathbf{H}_{i}^{\text{eff}} +\frac{\alpha}{M_s} \mathbf{S}_{i} \times \frac{\partial \mathbf{S}_{i}}{\partial t}\\
& - u_{i} \mathbf{S}_{i} \times \mathbf{S}_{i} \times \frac{\partial \mathbf{S}_{i}}{\partial x} - \beta u_{i} \mathbf{S}_{i} \times \frac{\partial \mathbf{S}_{i}}{\partial x} \\
\label{llg}
\end{aligned}
\end{equation}

In this equation, the first term represents the precession of the atomistic magnetic moments,  the second is the damping term, followed by the spin transfer torque (STT) term introduced by Sloncewski \cite{Slonczewski1996, Slonczewski2002}. The STT term describes the magnetization dynamics under the influence of a spin-polarized current \rjb{along the $x$-direction}. The constant $\alpha$ is the Gilbert damping parameter, $\gamma$ is the gyromagnetic ratio, and $\beta$ is the adiabaticity constant. The term $\mathbf{H}_{i}^{\text{eff}}$ is the effective field, which is defined as the partial derivative of the Hamiltonian \eqref{hamiltonian} with respect to the magnetic moment at each site \zw{$\boldsymbol{S}{_i}$}, ie., 

\begin{equation}
    \mathbf{H}^{\text{eff}}_i = - \frac{ \partial \mathscr{H}}{\partial \mathbf {S}_{i}} 
    \label{eq:dhdm}
\end{equation}

\noindent The term $u_i$ is the STT coefficient, given by 
\begin{equation}
 u_i=j_e P g \mu_B /\left(2 e M_s\right)
\end{equation}
where $j_e$ is the current density \rjb{along the $x$-direction}, $P$ is the spin polarization, and $M_s$ is the saturation magnetization.  For the simulations, we are using a simple cubic lattice with localised magnetic moments at the atoms. A periodic boundary condition is applied along the $xy$-plane, mimicking the physics of an infinite plane to avoid any boundary effects, especially during the current induced dynamics of skyrmions. \rj{ The Monte-Carlo simulation typically has steps in the order of $10^6$, depending on the complexity of the configuration. These steps are divided into different phases with decreasing temperature, chosen to establish that the final state obtained represents the global minimum and the noise in the average of the observable is minimized}. This is confirmed by monitoring the variation in the average energy of the system and ensuring that there is no significant change after each step.  

Once the ground states are obtained using MC simulation, the dynamics    are studied with atomistic spin dynamics calculations. The spin dynamics simulations are conducted at $\unit[0]{K}$ temperature with a time-step \zw{$dt = 10^{-16}$} seconds. To study the current-induced dynamics of the skyrmions, a spin-polarized current is applied along the +$x$-direction and the evolution of the magnetic moment at each point is recorded at appropriate time intervals. Simulations are repeated by varying the values of current strength $j_e$ from $1.75 \times 10^{11}$ $A/m^2$ to $30 \times 10^{11}$ $A/m^2$ and adiabaticity term $\beta$ from 0.01 to 0.5. A value of 0.1 is typically used for the damping term $\alpha$. The number of simulation steps is appropriately chosen based on the velocity of the skyrmion's motion. This choice ensures that we can record a trajectory of sufficient length to calculate the average velocity accurately. From this, the skyrmion Hall angle is calculated as, $\theta_{SHE} = \tan^{-1}\left(\frac{V_{y}}{V_{x}}\right)$, where $V_y$ and $V_x$ are the vertical and horizontal components of the skyrmion velocity. 

\subsection{Exchange bias: model setup} 
\begin{figure}
    \includegraphics[width=0.4\textwidth]{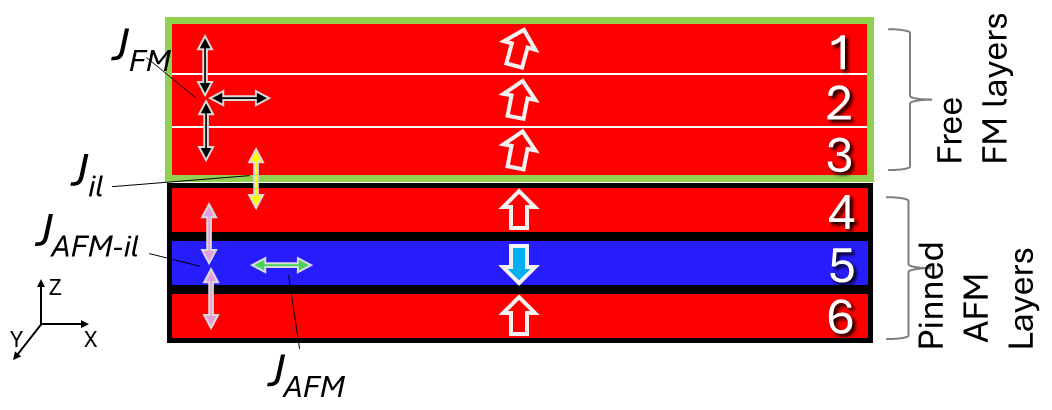}
    \caption{The model exchange bias setup, with different exchange interactions marked with arrows. Only three out of the seven AFM layers are shown.}
    \label{fig:jgraph}
\end{figure}

In the first stage, we investigated the exchange bias effect in AFM/FM multilayer structures {by simulating a hysteresis loop}. \rj{Our study included a comprehensive analysis of the dependence of the exchange bias field on various simulation parameters, such as the strength of the interlayer coupling ($J_{il}$) and the number of ferromagnetic layers.} \zw{ An illustrative diagram of \rj{an example setup which we used for our simulation} is given in Fig.~\ref{fig:jgraph}.} This model system is comprised of three ferromagnetic (FM) layers \zw{(labeled as layer 1, 2, 3 respectively)},  stacked over seven uncompensated antiferromagnetic (AFM) layers. In the AFM layers, the magnetization was pinned in the out-of-plane direction (along the Z direction), with a strong uniaxial anisotropy. This pinning is crucial for ensuring the stability of the magnetic orientations within these layers while the external field is varied. On the other hand, the FM layers' magnetization was not fixed and it was allowed to couple with the adjacent AFM layers through interlayer exchange coupling. This coupling plays a significant role in the exchange bias phenomenon observed at FM/AFM interfaces. The AFM/FM layers are stacked along the Z-axis direction. Atoms in the AFM layers are coupled to their nearest neighbors with \zw{in-plane} FM coupling ($J_{\text{AFM}}$) and with \zw{out-of-plane} AFM coupling ($J_{AFM-il}$).  The $4^{th}$ AFM layer is coupled to the $3^{rd}$ FM layer with an exchange strength given by $J_{il}$. The atoms in the FM layers are coupled to all neighbors with strength given by $J_{\mathrm{FM}}$. {The hysteresis loop is simulated by varying the external magnetic field from 1 T to -15 T and back to 1 T in steps of 0.5 T. \sout{A Monte Carlo (MC) simulation is conducted for each field step}. The first simulation occurs at a field step of 1 T, where the MC simulation begins from an initial state, in which the magnetization in the ferromagnetic (FM) layers is aligned along the +Z direction, while in the antiferromagnetic (AFM) layers, it alternates between -Z and +Z directions, as illustrated in Fig. \ref{fig:jgraph}. This simulation comprises four stages, with a total 250,000 steps, during which the temperature is decreased from 100 K to 10 K, and then to 1 K. This will yield the relaxed ground state at this specific field. This final state obtained here serves as the initial state for the subsequent \zw{field} stages of the hysteresis loop.}

\renewcommand{\arraystretch}{1.5} 
\begin{table}[h]
\caption{Material parameters used for the simulation of the model exchange bias setup}
\label{tab:exb}
\begin{tabular}{p{5cm}|p{1.3cm}|p{1.5cm}}
\toprule
\textbf{Parameter} & \textbf{Symbol} & \textbf{Value} \\
\toprule
Saturation Magnetization  & $M_{s}$ & $2.23 \mu_B$ \\
Exchange coupling in FM Layer & $J_{\mathrm{FM}}$ & 1.33 mRy \\ 
In-plane AFM coupling & $J_{\text{AFM}}$ & $J_{\mathrm{FM}}$ \\
Out of plane AFM coupling (along Z) & $J_{AFM-il}$ & $- J_{\mathrm{FM}}$ \\
Interlayer exchange coupling between FM/AFM Layers & $J_{il}$ & $0.025\, J_{\mathrm{FM}}$ to $0.4 \, J_{\mathrm{FM}}$ \\
Out of plane anisotropy in AFM Layers & $K_{AFM}$ & $-3\, J_{\mathrm{FM}}$ \\
Number of FM layers & $N_{FM}$ & 1 to 10 \\

\end{tabular}%
\label{tab:material_parameters}
\end{table}

We used ideal material parameters which are given in table \ref{tab:material_parameters}.  \sout{To keep the model simple, we are only considering the first neighbor exchange interaction} \rjb{ In our exchange bias simulation, we only considered the nearest-neighbor exchange interactions. This is based on typical behavior observed in the materials of interest, where nearest-neighbor coupling tends to be the dominant interaction and higher-order couplings, such as next-nearest neighbors, typically exhibit a magnitude approximately eight times smaller \cite{Loh2017, Gryt2019, Bornemann2019, Borisov2021, Borisov2022}}.  Since all the interactions are given in terms of $ J_{\mathrm{FM}}$, the model can be scaled up or down according to suit various material combinations.

\begin{figure}
    \includegraphics[width=0.49\textwidth]{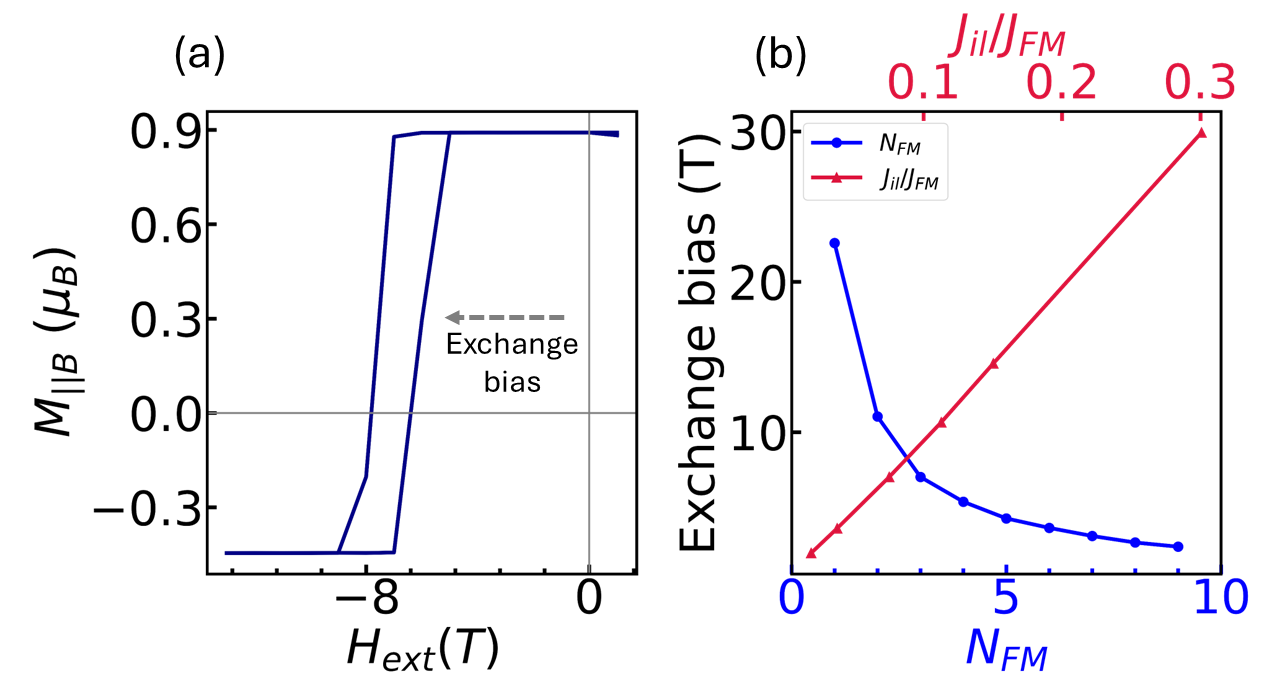}
    \caption{\rj{ Exchange bias effect in the model system. (a) Simulated hysteresis loop of the model system with three FM layers and $J_{il}/J_{\mathrm{FM}} = 0.07$. (b) Comparison of the exchange-bias strength with the inter-layer coupling strength $J_{il}$ (red) and the number of FM layers ($N_{FM}$) (blue). }}
    \label{fig:exb_graph}
\end{figure}

\subsection{Skyrmions in FM Layers: formation field}
\sout{Once the material parameters that could result in desired ranges of exchange bias field strengths are identified, we move to the second stage. The stability and characteristics of skyrmions in FM layers, with an emphasis on the effects of in-layer Dzyaloshinskii Moriya Interaction (DMI), were examined here.} To study the formation and stability of skyrmions, a ferromagnetic monolayer, with an in-plane DMI is studied. We did a range of large-scale simulations, systematically varying the strength of the ferromagnetic exchange strength from $0.1 J_{\mathrm{FM}}$ to $J_{\mathrm{FM}}$, the DMI strength  $D$ from $0.7 J_{\mathrm{FM}}$ to $J_{\mathrm{FM}}$, and the temperature from $\unit[0]{K}$ to $\unit[300]{K}$. The direction of the $\mathbf{D}$ was aligned parallel to the bond direction. These simulations identified the parameter space in which skyrmions could remain stable.

\subsection{Stability of skyrmions: GNEB method}\label{sec:GNEB}
 To assess the thermal stability of skyrmions, we explore the energy and transition mechanisms involved in skyrmion annihilation using the geodesic nudged-elastic band method (GNEB), \cite{Pavel2015}. {Magnetic skyrmions are separated from a topologically trivial colinear state by a finite energy barrier. To get an accurate description of their stability, and the mechanism involving their magnetic transitions the minimum energy path (MEP) involved and the corresponding activation energy should be studied. The activation energy is defined as the difference between the energy of the initial state and that of the highest saddle point (SP) along the MEP. The geodesic-nudged-elastic-band (GNEB) method is a technique that is adept at tracing the minimum energy paths over the complex, curved energy landscapes of such magnetic transitions. This involves the identification of the local minima at the initial ($\mathcal{G}^{int}$) and final ($\mathcal{G}^{fin}$) states and then connecting these two states by an initial path. This path is discretized to several points called the images. These images give the magnetic state of the system at these intermediate points and they are an approximation to a continuous path. The number of images in the energy path is kept at an optimum value so as to have a good resolution while still minimizing computational cost. \rjb{The initial path between $\mathcal{G}^{int}$ and $\mathcal{G}^{fin}$ is brought to the nearest minimum energy path, by means of an iterative algorithm that tries to minimize the force in the tangent space, which is calculated similarly to Eq. \ref{eq:dhdm}}. Once the MEP is identified, the energy barrier or the activation energy can be calculated as
 \begin{equation}
    E_{ac} = E(\mathcal{G}^{SP})-E(\mathcal{G}^{int}) 
\end{equation}
Where $\mathcal{G}^{SP}$ is the saddle point. This $E_{ac}$ is a measure of the thermal stability of the skyrmion. It relates to the lifetime of the skyrmion in the form of an Arrhenius rule \cite{Bessarab2018, kramers1940brownian, bessarab2012harmonic}. A detailed description of the theory and the algorithm of this technique can be found in Ref. \cite{Pavel2015}. 

\section{Results and discussion}

\subsection{Exchange bias} \label{sec_Exb}
\rjb{ The exchange bias effect observed in FM/AFM multilayer structures is characterized by an asymmetric magnetization reversal and a consequent shift in the coercive field. This is caused by the presence of a unidirectional anisotropy at the FM/AFM interface due to interfacial coupling, which acts as an additional field term during magnetization reversal \cite{Billoni2011, Ohlag2003, Stamps2000}. The shift in the coercive field gives the strength of the exchange bias field $H_{\text{Exb}}$}.  The hysteresis loop obtained for the model setup with three FM layers and an FM-AFM interlayer coupling strength ($J_{il}$) of 0.075 $J_{\mathrm{FM}}$ is shown in Figure \ref{fig:exb_graph}. The loop is shifted towards the negative X-axis, signifying the presence of an exchange bias field of 7.7 T. This simulation is repeated, systematically varying the interlayer exchange coupling $J_{il}$ from 0.01 to 0.4 $J_{\mathrm{FM}}$ and by increasing the number of FM layers from 1 to 10. The results are shown in Fig. \ref{fig:exb_graph}. We observed that the exchange bias field strength was directly proportional to the interlayer coupling strength (Fig.~\ref{fig:exb_graph}).  \rjb{This result underscores the fundamental role of inter-layer coupling in generating the exchange bias effect and agrees with the existing theoretical models \cite{Meiklejohn1962, Mauri1987} and previously published simulation results, e.g., in Ref. \cite{Oscar2005, Garca2010, Billoni2011}. In our model, the AFM spins at the interface are pinned along the Z direction with strong uniaxial anisotropy while the free FM spins are coupled to these pinned spins by the interlayer coupling $J_{il}$. Thus the exchange field experienced at the interface is directly proportional to $J_{il}$}. Conversely,  we observed an inverse relationship between the exchange bias field and the number of ferromagnetic (FM) layers, as shown in figure.~\ref{fig:exb_graph} (b). This phenomenon can be attributed to the specific characteristics of our model, where only first-neighbor coupling occurs between atoms at the FM/AFM interface. Consequently, the exchange bias is predominantly an interfacial effect, confined primarily to the first FM layer adjacent to it. The influence of the exchange bias extends to the above-lying FM layers via the $J_{ij}$ coupling. As the number of layers increases, the intensity of the exchange bias effect diminishes in the layers farther from the interface, explaining the observed inverse relationship \cite{exb_layer_thickness}. Based on these results, parameter combinations that result in desired values of exchange bias field strength could be identified. These findings are particularly relevant for the next stage, where we stabilize skyrmions with exchange bias.

\subsection{Skyrmion in Ferromagnetic layers}

\rjb{In the study of skyrmions within FM/AFM multi-layers, we introduce an in-plane Dzyaloshinskii-Moriya interaction (DMI) in the ferromagnetic layer. This incorporation of DMI in the FM layer may influence the exchange bias field}. Consequently, the field impacting a FM layer with DMI might vary from that affecting a FM layer without it, assuming all other conditions are identical. The discrepancy in the exchange bias field is challenging to measure accurately due to the difficulty in modeling a collinear state at zero fields caused by the presence of DMI. This complication hinders identifying the hystereses loop and, consequently, the coercive field. Although there have been theoretical explorations of the exchange bias field strength at FM/AFM interfaces, \cite{Yanes2013}, our simulations suggest that the variation caused by DMI in our setup is likely minor. Therefore, for simplicity, we overlook any potential differences in the exchange bias field due to the effect of DMI in the FM layer.

\sout{The material parameters are chosen based on the insights from the previous results.} \rjb{To study the formation of skyrmions,  with the Monte Carlo simulation approach}, we initiate the magnetization process from a state where each site's magnetization is randomly oriented, simulating the physical case of thermal annealing. \rj{From this initial state the magnetization is relaxed to a zero-field state.} In the absence of an external field, the magnetization naturally settles into a labyrinthine domain state, as shown in Fig.~\ref{fig:phase_space}. This state is distinguished by its intricate, tangled pattern. The width of these domains typically aligns with the size of skyrmions found in the material \cite{Wang2018}. \rj{To this state, an out-of-plane field is applied}. When we gradually increase the out-of-plane magnetic field strength, these domain structures begin to evolve. Domains aligned with the external field expand, while those in opposition contract. Increasing the field further leads to the formation of skyrmions within the layer, as depicted in Fig.~\ref{fig:phase_space}. Continuing to raise the field strength causes the skyrmions to reduce in size until, at a certain high field strength, they are annihilated, resulting in a saturated or collinear magnetic state. This field strength at which a skyrmion is formed: the formation field, which is crucial for the creation of a skyrmion, results from the interplay of the exchange coupling strength $J$ and the Dzyaloshinskii-Moriya interaction strength $D$. Consequently, the formation field is influenced by the $D/J$ ratio inherent to the material. \rj{To investigate this, we conducted a series of simulations varying the $D$ and $J$ values systematically. In each simulation, we identified the formation field as the strength of the field at which the skyrmions are formed when transitioning from a labyrinth state. The results are illustrated in Fig. ~\ref{fig:formationfield}. It is observed that the formation field is directly proportional to the in-plane exchange coupling $J$ and the $D/J$ ratio. Based on the values of $J$ and $D$, the formation field varied considerably from $\unit[0.7]{T}$ to $\unit[85]{T}$.\sout{ From this table, the field strength required to stabilize the skyrmions at various values of $J$ and $D$ can be identified.} \rjb{The material parameters associated with different exchange bias field strengths were already identified based on the studies conducted in section \ref{sec_Exb}. Thus, combining these two results, } exchange bias-stabilized skyrmions are simulated by stacking the FM layers with in-plane DMI on top of a pinned AFM layer. The simulated skyrmion and an illustration of the setup are given in Fig.~\ref{fig:skyr_setup}. The material parameters used in this setup are given in Table \ref{tab:parameters_skyrmion}. These values are carefully chosen by comparing the previous results so that the exchange bias field arising from the FM/AFM interface exactly matches the formation field of the skyrmions in the FM layer.}

\renewcommand{\arraystretch}{1.5} 
\begin{table}[h]
\caption{Material parameters used for the simulation of skyrmions}
\label{tab:skyr}
\begin{tabular}{p{5cm}|p{1cm}|p{1.5cm}}
\toprule
\textbf{Parameter} & \textbf{Symbol} & \textbf{Value} \\
\toprule

Exchange coupling in FM layer & $J_{\text{FM}}$ & $1.3 \, \text{mRy}$ \\
        In-plane DMI strength & $D$ & $0.1 J_{\mathrm{FM}}$ \\
        In-plane AFM coupling & $J_{\text{AFM}}$ & $J_{\text{FM}}$ \\
        Out-of-plane AFM coupling & $J_{\text{AFM-il}}$ & $-J_{\text{FM}}$ \\
        Interlayer exchange coupling between FM/AFM Layers & $J_{\text{il}} (\text{N1})$ & $0.011 \, J_{\text{FM}}$ \\

\end{tabular}%
\label{tab:parameters_skyrmion}
\end{table}

\begin{figure}
    \includegraphics[width=0.45\textwidth]{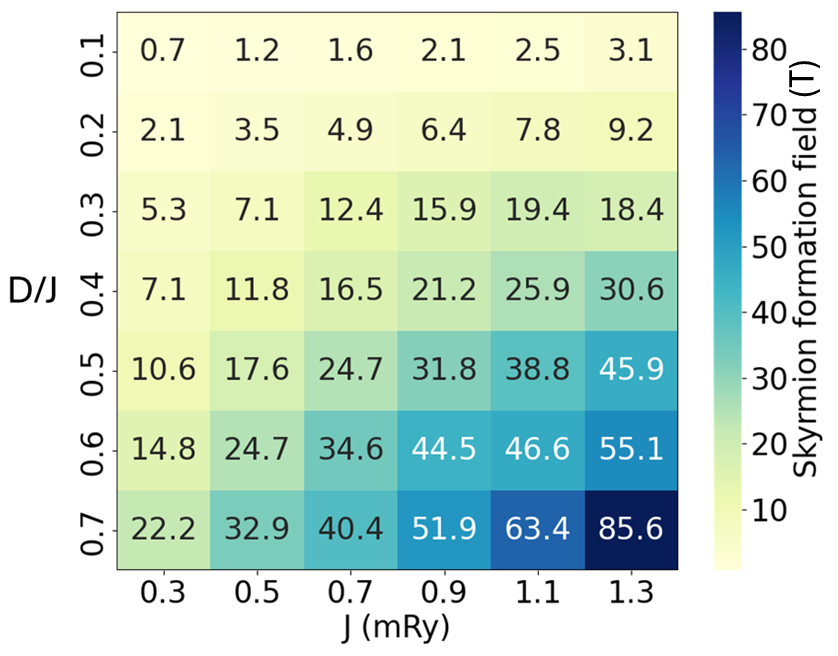}
    \caption{Skyrmion formation field, at various values of exchange strength $J$ and $D$. The formation field is directly proportional to $J$ and $D/J$ ratio.}
    \label{fig:formationfield}
\end{figure}

\begin{figure}
    \includegraphics[width=0.45\textwidth]{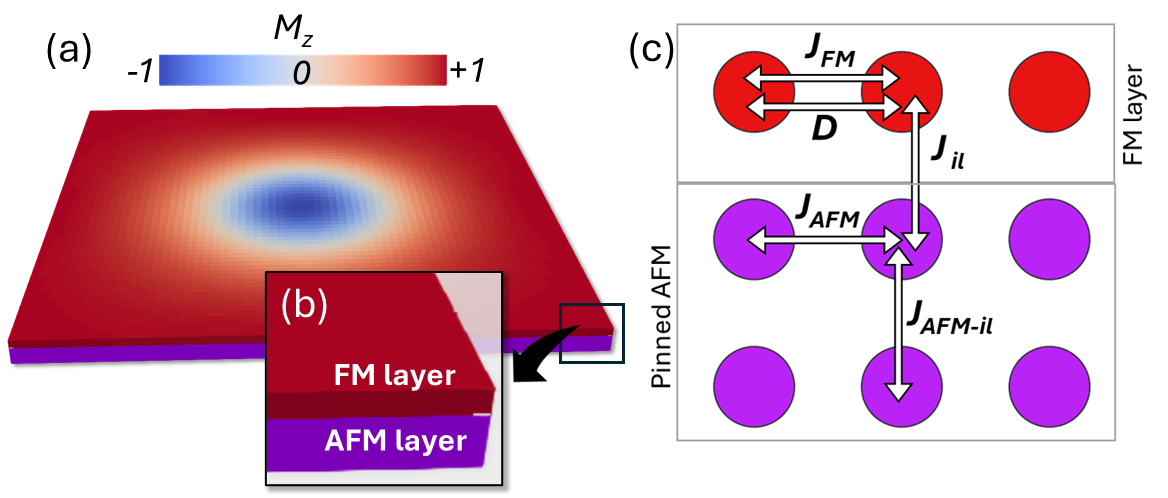}
    \caption{ \rj{(a) Zero-field skyrmion stabilized by the exchange bias field in the FM/AFM multilayer setup. \rjc{The colors describes the out of plane (Z) component of the Magnetization as indicated in the legend. Red color indicates a large spin up component perpendicular to the surface, blue color indicates a large spin down component perpendicular to the surface and a white color indicates that the spin is parallel to the surface. The bottom purple layer represents the AFM layer } (b) Zoomed view showing the top FM and bottom AFM layer (c) Inter-atomic exchange ($J$) and DMI coupling ($D$) in this setup.}}
    \label{fig:skyr_setup}
\end{figure}

\begin{figure}
    \includegraphics[width=0.5\textwidth]{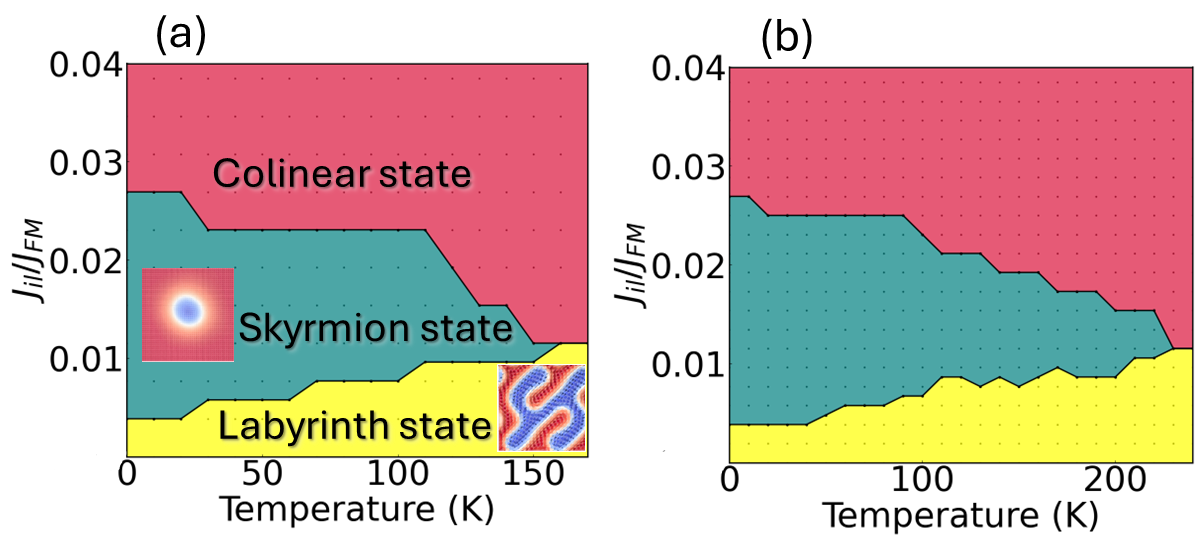}
    \caption{\rjb{The stability of skyrmions with respect to temperature and the interlayer coupling $J_{il}$ in (a) an FM monolayer under exchange bias (Fig. ~\ref{fig:skyr_setup}), and (b) an AFM-coupled double layer under exchange bias (Fig. ~\ref{fig:conduit}(d)). The yellow region is the parameter space in which the labyrinth domain states are observed, and the red part is the area in which the system exists in a colinear state. Stable skyrmions are observed in the green area.
    }}
    \label{fig:phase_space}
\end{figure}

A key metric for assessing the stability of skyrmion states within this setup is the range of field strengths within which the skyrmion maintains its stability. To determine this, one begins by establishing a stable skyrmion as the initial state. Subsequently, the exchange bias field is methodically incremented and decremented by varying the interlayer coupling strength $J_{il}$. This process identifies the upper and lower limits of the magnetic field strengths at which the skyrmion ceases to be stable. The span of the field strengths between these two limits offers a quantifiable indication of the  skyrmion's stability against field fluctuations, in the given material. To study the effect of temperature on this, the process is repeated by varying the temperature. From this, the temperature-field space inside which the skyrmion is stable, can be identified. The results of these calculations are given in Fig.~\ref{fig:phase_space}(a). Due to the FM layers being stacked atop a pinned AFM layer and the exchange bias acting as the field source, the field strength is quantified in terms of the interlayer exchange coupling $J_{il}$.  In lower field regions, the labyrinth phase is predominant, whereas at higher field strengths, the system transitions into a collinear state. Notably, there exists a well-defined range of field strengths, sandwiched between these two phases, where the skyrmions exhibit stability. It was observed that as the temperature increases, this stability field range for the skyrmions progressively narrows. Beyond a critical temperature threshold of 160K,  it ceases to be a field range where the skyrmion can maintain stability. \rjb{In addition to the single FM monolayer, we investigated a setup involving two AFM-coupled FM layers (Fig. \ref{fig:phase_space} (b)), sandwiched between two pinned AFM layers. This configuration is significant later in our discussion and more details about this setup are given in section ~\ref{sec:conduit}. We observed qualitatively similar results as the monolayer, however, the double layer configuration demonstrated enhanced temperature stability, persisting up to $\unit[230]{K}$, compared to  $\unit[160]{K}$ the monolayer.}

\subsection{Stability: annihilation and activation energy}
To better understand the thermal stability of the skyrmions in these structures, their annihilation mechanism in an FM monolayer and an FM/AFM multilayer were studied in detail with GNEB, as explained in the methodology section \ref{sec:GNEB}. \rj{For this, an initial path discretized into $20$ images was created, where the an isolated skyrmion is chosen as the initial state and a collinear FM state is chosen as the final state. The GNEB calculation \rjb{gives the minimum energy path (MEP) with respect to the reaction coordinate ( see Fig. \ref{fig:gneb}) and the corresponding magnetization dynamics.}}  \sout{The minimum energy path of this magnetic transition in terms of the reaction coordinate is given in \ref{fig:gneb}.} \rj{Reaction coordinate is a dimensionless number representing the progress of the magnetic transition and it is defined as a variation along the path connecting the initial and final states. It is normalized so that it is equal to zero at the initial state and one at the final state}. The energy of the system at each point along the path is plotted on the Y-axis, taking the energy of the skyrmion state as zero. In both these configurations, the annihilation of the skyrmion follows a similar \rjb{path.} \sout{pattern.} Initially, there is a progressive shrinking of the skyrmion's core, \rj{the energy of the skyrmion is gradually increased during this phase. The energy reaches its peak value when the skyrmion is in the smallest size. This is the saddle point (SP)}. The SP is followed by \rj{an abrupt} reversal of the outer region \rj{leading to the destruction of the skyrmion leaving behind a non-colinear spin structure at the core. These spins undergo a smooth rotation, while the energy of the system is gradually decreased,} eventually leading to a collinear state.  This observed collapse mechanism is consistent with \rjb{previous simulation studies \cite{Pavel2017_Skyr_collapse}  and } recent experimental observations \cite{Muckel2021}.  In the energy landscape, the skyrmion state represents a metastable local minima state and the global energy minimum is given by a colinear magnetic state. \sout{ Between these two points, there is the saddle point given as a peak in the energy curve. The height of this peak from the zero level is the activation energy ($E_{ac}$) of the skyrmion, which is a direct measure of the thermal stability and the mean lifetime of a skyrmion.}\rjb{ The activation energy ($E_{ac}$)  for the transition from the skyrmion to the collinear state is given by the height of the saddle point. This $E_{ac}$ is a direct measure of the thermal stability and the mean lifetime of a skyrmion}. 

When comparing the activation energies required for the annihilation process in two distinct setups, we notice a significant 40\% increase when the skyrmion is stabilized by an exchange bias field. This points to a notably higher stability of the skyrmion in the exchange bias setup compared to a monolayer configuration.  The reason for this increased stability is linked to the formation of a skyrmion ``imprint'' \cite{Wu2011, SalazarAlvarez2009, Baltz2018_rev, Rana2021} in the neighboring layer, a process aided by exchange interactions. Due to the strong exchange coupling at the interface, the magnetizations in the antiferromagnetic (AFM) and ferromagnetic (FM) layers tend to align with one another. As a result, any magnetization texture that forms in the FM layer is mirrored in the AFM layer, leading to the creation of an ``imprint'' of the skyrmion in the lower layer, hence increasing the stability of the skyrmion. \rj{This is further verified by comparing the activation energy of the skyrmion to a case where there is a high uniaxial anisotropy in the AFM layers, restricting the formation of these imprints. In this scenario, the activation energy was the same as that of the case stabilized by an external field.} The details of the theory and the mechanism that governs this interaction have been discussed in \cite{Zelent2023}.

\begin{figure}
    \includegraphics[width=0.5\textwidth]{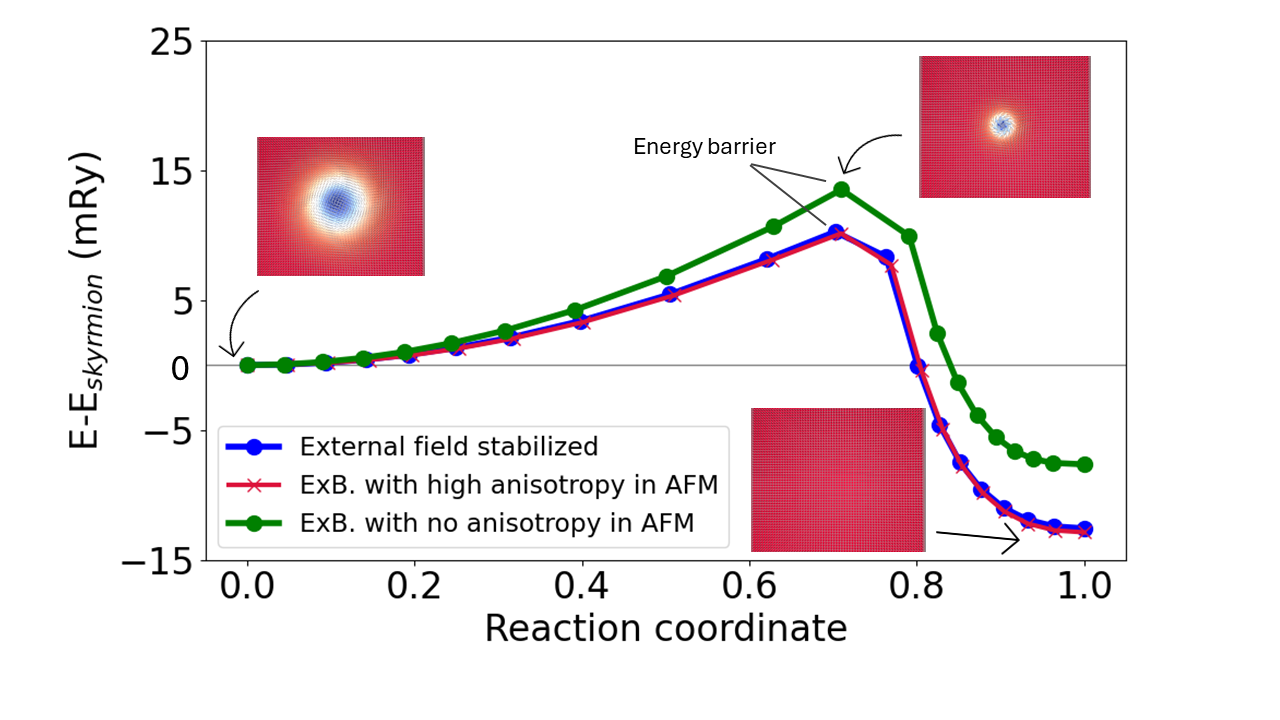}
    \caption{\rj{Minimum energy path illustrating the mechanism and energy barrier involved in the annihilation of the skyrmion. Each image depicts the magnetization state at a specific point along the reaction path. The blue curve represents the minimum energy path (MEP) for the skyrmion stabilized by an external field. The top green curve illustrates the MEP for the skyrmion in the absence of anisotropy in the AFM layers. The increase in $E_{ac}$ is clearly visible. The red line corresponds to the MEP for a skyrmion stabilized by an exchange bias in the presence of high anisotropy in the antiferromagnetic (AFM) layers. } }
    \label{fig:gneb}
\end{figure}

\subsection{Dynamics}
Having established a deep understanding of the stability of the skyrmions in different configurations and material parameters, we can proceed to look at their current-induced dynamics. In the first part, we investigated the dynamics of isolated skyrmions stabilized by exchange bias in monolayers, focusing on skyrmions of up and down polarity, \rjb{but with the same chirality}. \rj{We use the following values for the simulation parameters: exchange coupling within the FM layer  $ J_{\mathrm{FM}}$, DMI strength within the FM layer $D = 0.1 J_{\mathrm{FM}}$, inter-layer exchange coupling between the FM and AFM layers, $J^{1}_{il} = 0.01 J_{\mathrm{FM}}$. These values are chosen based on the previous results so that we can have skyrmions with high stability, small size, and realistic values of the exchange bias field. A smaller skyrmion will let us use a smaller simulation box, thus reducing the computational cost. Simulations were repeated at different values of current strength $j_e$, $\alpha$, and  $\beta$. From these series of calculations, the velocity and $\theta_{SHE}$ are calculated,} \rjb{which are given in Fig. ~\ref{fig:hall_angle}(a)}.

\rjb{ In this scenario, a skyrmion with a polarity of -1 demonstrates negative Hall angles when $\beta$ is less than $\alpha$, set at 0.1. Consequently, it drifts towards the lower section of the racetrack due to the Hall effect. When $\beta$ equals $\alpha$, the skyrmion shows no Hall effect, maintaining a linear trajectory devoid of any vertical displacement. As $\beta$ surpasses $\alpha$, the skyrmion exhibits positive Hall angles. The magnitude of this Hall angle increases linearly with $\beta$, escalating by a factor of 40 as $\beta$ progresses from 0.1 to 1, while the skyrmion's velocity experiences only a slight rise from 280 to 300 m/s (see Fig. ~\ref{fig:hall_angle}(b)). Our investigation affirms the theoretical projections \cite{Brearton2021, Tomasello2018, Jiang2016, Iwasaki2013},  that skyrmions of opposing polarities present Hall angles that are equal in magnitude yet opposite in sign. Thus 
replacing a skyrmion with opposite polarity but the same chirality results in Hall angles at different $\beta$ values that are equal in size but opposite in direction. }

Initially, we hypothesized that pairing two such skyrmions with opposite polarities \cite{Zhang2016}, each exhibiting equal and opposite Hall angles individually, and coupling them through an antiferromagnetic (AFM) interaction would lead to their movement without any Hall angles.} To test this, we created a bilayer arrangement of skyrmions, \textcolor{black}{as depicted in Fig.~\ref{fig:exb_ds}}. This system, referred to as setup 1, consists of four layers. The first ferromagnetic layer (layer 1) is magnetized in the -Z direction and is coupled to the second layer (layer 2) through a first neighbor antiferromagnetic coupling ($J^{1}_{sk}$). Layer 2 is further coupled to a pinned antiferromagnetic layer (layer 3) via nearest neighbor ferromagnetic coupling ($J^{1}_{il}$). \rjb{In contrast to the previous monolayer setup, this configuration additionally incorporates a next-nearest-neighbor ($N2$)  AFM coupling ($J^{2}_{il}$) between layer 1 and layer 3.} \sout{while layer 1 connects to layer 3 through a second nearest neighbor ($N2$) antiferromagnetic coupling ($J^{2}_{il}$).} For the simulation, we used the following values: $J^{1}_{sk} =  J_{\mathrm{FM}}$, $J^{1}_{il} = 0.01 J_{\mathrm{FM}}$, $J^{2}_{il} = -0.005 J_{\mathrm{FM}}$. \sout{These values were selected such that the exchange bias field formed due to these exchange strengths equals the skyrmion stabilization field.} In this setup, layers 1 and 2 each host a skyrmion. These skyrmions, with opposite polarities, are antiferromagnetically coupled to each other, effectively functioning as a unified entity. This unique arrangement of coupled skyrmions with contrasting polarities and their stabilization through exchange interactions creates a novel platform.  \sout{It enables the exploration of magnetic interactions and their effects on skyrmion dynamics, particularly regarding the mitigation of the skyrmion Hall effect. The current-induced motion of this coupled skyrmion is studied under different values of $\alpha$, $\beta$, and current strength $j_e$ are studied and the skyrmion Hall angle is measured.}\rjb{The figure depicted in Fig. \ref{fig:hall_angle} illustrates the velocity and Hall angle of the coupled skyrmion at various values of $\beta$, with $\alpha$ fixed constant at 0.1. Here, the velocity of the skyrmion shows a linear increase with $\beta$. Within the range $\beta < 0.1$, the coupled skyrmion moves at a velocity slower than that of the isolated monolayer skyrmions. However, as $\beta$ surpasses 0.1, the coupled skyrmions demonstrate velocities significantly faster than those of the isolated skyrmions, reaching speeds of up to 2.5 Km/s when $\beta = 1$. }\rjb{Nevertheless, we observed an unusual behavior regarding the skyrmion Hall angle.} \rj{In the range $\beta > 0.4$, the skyrmion Hall angle, $\theta_{SHE}$ of the coupled skyrmions drops below the individual  $\theta_{SHE}$  of the monolayer skyrmions. The coupled skyrmions move with very low Hall angles ($\theta_{SHE} < 10\degree$), despite the isolated skyrmions showing very high Hall angles ($> 20 \degree$) in this region}. However, when $\beta < 0.3$ the coupled skyrmion exhibits a Hall angle significantly greater than the sum of the individual skyrmions' angles, tending towards $90 \degree$. This unexpected result can be attributed to the asymmetry in the exchange fields experienced by FM layers 1 and 2. Layer 1 is coupled to the AFM layer via second-neighbor coupling ($N2$), whereas layer 2 experiences first-nearest-neighbor ($N1$) coupling. This disparity leads to unequal exchange fields for the top and bottom skyrmions, resulting in their size deformation and, consequently, a heightened Hall effect in the coupled state.

    To address the observed anomaly, we propose the creation of an artificial setup (Fig.~\ref{fig:exb_ds}, Setup 2) that ensures uniformity in the exchange field experienced by both layers. In this revised configuration, both the top and bottom ferromagnetic (FM) layers are coupled to the underlying antiferromagnetic (AFM) layers using second-nearest neighbor ($N2$) coupling, \rj{and there is no coupling between layers 2 and 3}. Such a design makes sure that the exchange field across both layers is equal, thereby eliminating the asymmetry observed in the original setup. In this artificial setup, the uniform coupling results in the coupled skyrmion moving with a zero net Hall effect across the entire range of ($\beta$). This outcome is significant as it demonstrates the ability to control and negate the skyrmion Hall effect through careful design of the magnetic layer coupling. By ensuring that both layers are subjected to the same magnetic influences, the coupled skyrmion behaves consistently, regardless of the value of $\beta$.

\begin{figure}
    \includegraphics[width=0.5\textwidth]{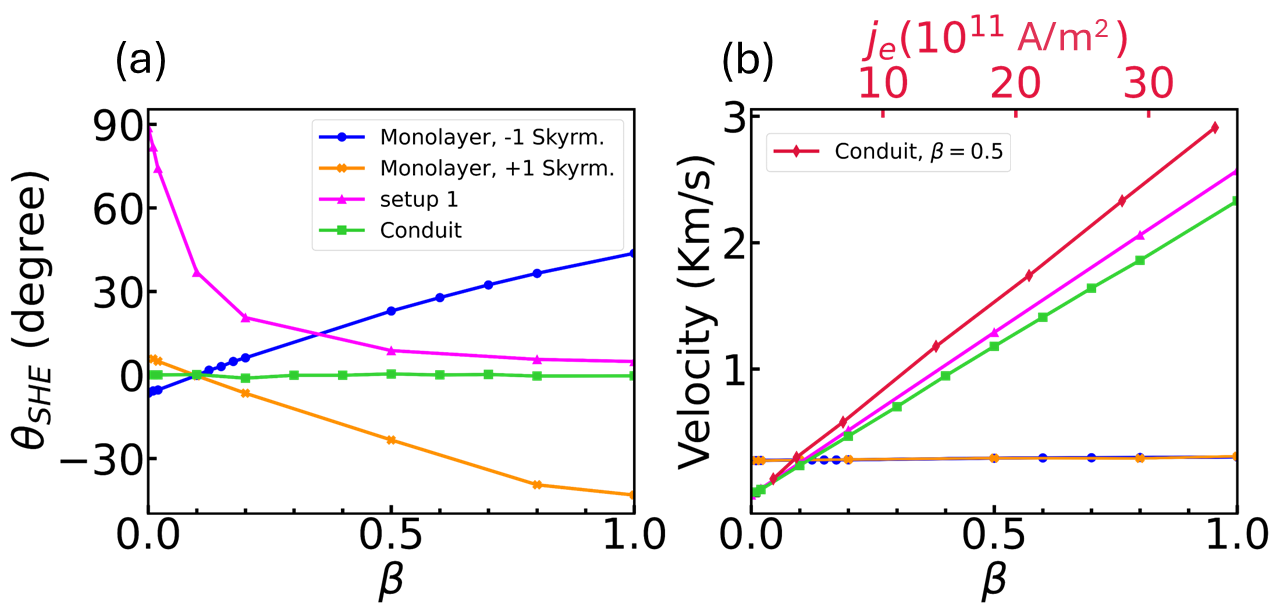}
    \caption{\rjb{(a) Variation in the skyrmion Hall angle of the isolated skyrmions with +1 and -1 polarity in the monolayer and coupled skyrmions in setup 1 and the conduit with different values of $\beta$ when $\alpha =0.1$ and $j_e=14 \times 10^{11}$\,A/m$^2$.
    (b) Horizontal velocity of these skyrmions with $\beta$. The red line shows the velocity of the coupled skyrmion in the conduit with different current densities ($j_e$) keeping $\alpha=0.1$ and $\beta=0.5$. }}
    \label{fig:hall_angle}
\end{figure}

\begin{figure*}
    \includegraphics[width=\textwidth]{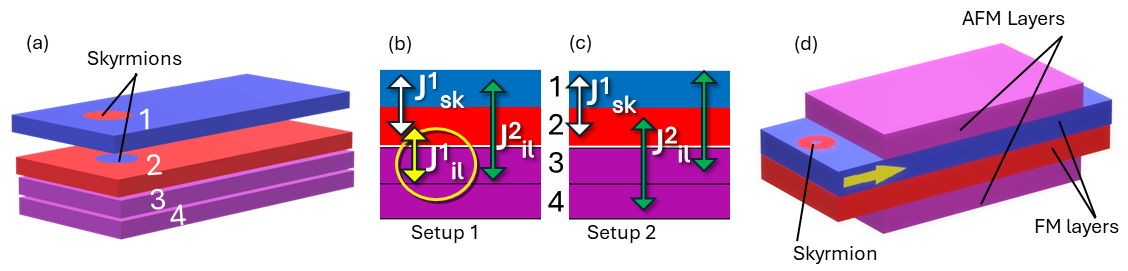}
    \caption{\rj{(a) Model system for studying skyrmion dynamics. Layers 1 and 2 are the skyrmion-hosting FM layers, while layers 3 and 4 are the pinned AFM layers. Skyrmions of up/down polarity in the two layers are stabilized by exchange bias from the bottom AFM layer \label{fig:exb_ds} (b) Exchange interaction in setup 1. The FM layers 1 and 2 are coupled to the AFM layers with a first nearest neighbor (N1) and second nearest neighbor (N2) coupling.   (c) In Setup 2, both FM layers are coupled to the AFM layers with N2 coupling. (d) Exchange bias stabilized skyrmion conduit. The coupled skyrmion is hosted in the middle FM layers. Exchange bias arising from the two AFM layers keeps the skyrmion stable while the AFM coupling between the first and second FM layers cancels the skyrmion Hall effect.} \rjc{The arrows in (b) and (c) represent the exchange coupling between the layers. The white arrow represents the nearest neighbor (N1) AFM coupling, while the yellow arrow shows the nearest neighbor FM coupling, and the green arrow gives the next-nearest-neighbor (N2) coupling. }
    }
    \label{fig:conduit}
\end{figure*}

\subsection{Exchange-bias stabilized skyrmion conduit} \label{sec:conduit}

Based on the insights gained from these studies, we propose the design for a skyrmion transport device capable of hosting stable skyrmions at zero magnetic fields and transporting them at high velocities while completely negating the Hall effect. This device is structured by stacking two skyrmion-hosting ferromagnetic (FM) layers on top of one another, with these layers being coupled via nearest-neighbor antiferromagnetic (AFM) interactions. The entire arrangement is then enclosed between two additional antiferromagnetic layers, as depicted in Fig.~\ref{fig:conduit}. In this configuration, the first FM layer (FM layer 1) is connected to the second FM layer (FM layer 2) through AFM coupling and to the bottom AFM layer via FM coupling. The second FM layer (FM layer 2) is also coupled to the top AFM layer with FM coupling. The exchange bias field arising from the coupling to the AFM layers keeps the skyrmions stable and as the two skyrmions within the FM layers are coupled in an antiferromagnetic manner, their respective Hall effects counteract each other. This reciprocal cancellation ensures that the skyrmions can be transported at high velocities without experiencing any Hall effect (Fig. ~\ref{fig:hall_angle}).  \rj{The velocity of the coupled skyrmion in the conduit at different values of $\beta$ and $j_e$, while fixing the value of $\alpha$ to 0.1 are given in Fig. \ref{fig:hall_angle}(b)}.\rjb{The skyrmion velocity increased linearly with $j_e$ and $\beta$ and we observed} high velocities up to 3 \,\text{Km/s}.  \rjb{In comparing the velocities of coupled skyrmions in the conduit versus those in setup 1, it's evident that the velocity of the skyrmion in the conduit experiences only a marginal decrease, while successfully circumventing the Hall effect entirely. The high velocity reported here is at least six times faster than the skyrmion velocities generally reported before \cite{Lai2017, Kolesnikov2018vel, Gobel2019}.} Thus,  such a device, leveraging the principles of magnetic layer coupling and exchange bias stabilization, offers a promising avenue for efficient and controlled skyrmion transport in spintronic applications.

\subsection{Skyrmions in antiferromagnetic layers}
In addition to our primary study on skyrmions in FM layers stabilized by exchange bias from AFM layers, we extended our research to include skyrmions in two distinct antiferromagnetic (AFM) structures: A-type AFM structures stabilized by exchange coupling from a pinned FM layer, and in a G-Type AFM interface\cite{Baltz2019}.

 \subsection*{Skyrmions in A-type AFM Structures Stabilized by Pinned FM Layers}
\begin{figure}
    \centering
    \includegraphics[width=0.4\textwidth]{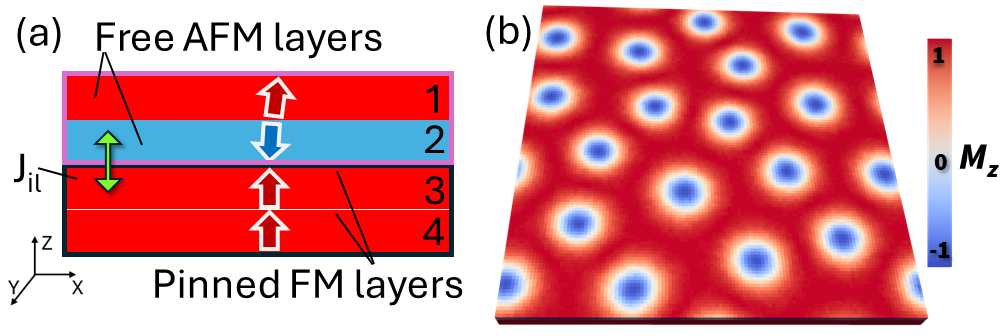}
    \caption{(a) Model setup for the simulation of skyrmions in an AFM layer. \rjc{ Layers 1 and 2 are the AFM layers and layers 2 and 3 are the pinned FM layers. The arrows represent the direction of magnetization of the layer.}  (b) Skyrmion lattice formed in the AFM layer. }
    \label{fig:reverse_setup}
\end{figure}

The simulation setup for this investigation mirrors our primary model system. It consists of three pinned FM layers which are coupled to two A-type AFM layers by an antiferromagnetic coupling ($J_{il} < 0$). The distinguishing feature in this setup is the location of the Dzyaloshinskii-Moriya Interaction, which is present only in the AFM layers. We adopted the same material parameters as used in our primary study to maintain consistency and comparability of results. An illustration of the model is given in Fig.~\ref{fig:reverse_setup} (a). Here, the Dzyaloshinskii Moriya  interaction is confined to the antiferromagnetic (AFM) layers, and the magnetization within the  FM layers is fixed along the Z-axis, with a strong uniaxial anisotropy. This setup leads to the formation of magnetic skyrmions in the AFM layers as shown in Fig.~\ref{fig:reverse_setup}(b).  The exchange coupling from the stationary ferromagnetic layers stabilize these skyrmions. Given that the exchange interaction and other relevant parameters closely resemble those in earlier configurations, the skyrmions within the AFM layers exhibit current-driven dynamics akin to those outlined in section III-D, \rj{in the sense that they move with equal and opposite Hall angles when isolated and exhibit the anomaly in the low $\beta$ region when coupled together.}

\subsection*{Skyrmions in G-type AFM Interface}\label{sec_gtypefam}
The G-type AFM setup presents a more complex scenario. It can be conceptualized as two interlaced FM sublattices \cite{Baltz2019}, akin to a chessboard pattern, with atoms in each lattice coupled through distinct exchange interactions.  The second nearest neighbor atoms, belonging to the same sublattice are coupled with ferromagnetic exchange coupling ($J^2$), while the next nearest atoms in the opposing lattice are linked via antiferromagnetic coupling ($J^1$). Additionally, this structure incorporates an in-plane DMI. The magnetizations within these lattices are pinned in the out-of-plane or Z direction, with the first lattice aligning along the +Z direction and the second lattice along the -Z direction. The material parameters used for the simulation of skyrmion in this G-type AFM structure are given in table \ref{tab:gtype-afm}. These parameters were chosen to optimize the conditions under which skyrmions could exist stably within the G-type AFM structure.

\begingroup

\renewcommand{\arraystretch}{1.5} 
\begin{table}[h]
\caption{Material parameters for \rjc{the simulation of stable skyrmions in a G-type AFM interface ~\ref{sec_gtypefam}. }}
\label{tab:material_parameters_gtype}
\begin{tabular}{p{5cm}|p{1cm}|p{1.5cm}}
\toprule
\textbf{Parameter} & \textbf{Symbol} & \textbf{Value} \\
\toprule
Exchange coupling between second nearest neighbors in the FM Layer & $J_{\mathrm{FM}}^{2}$ & $J_{\mathrm{FM}}$ \\
Exchange coupling between first nearest neighbors from opposing lattices in the AFM layers & $J_{\text{AFM}}^{1}$ & -2 $J_{\mathrm{FM}}$  \\
DMI strength & $D$ & 0.75 $J_{\text{FM}}$ \\
Uniaxial anisotropy constant & $K$ & 1.3 $J_{\mathrm{FM}}$ \\

\end{tabular}%
\label{tab:gtype-afm}
\end{table}

\endgroup

\begin{figure}
    \includegraphics[width=0.45\textwidth]{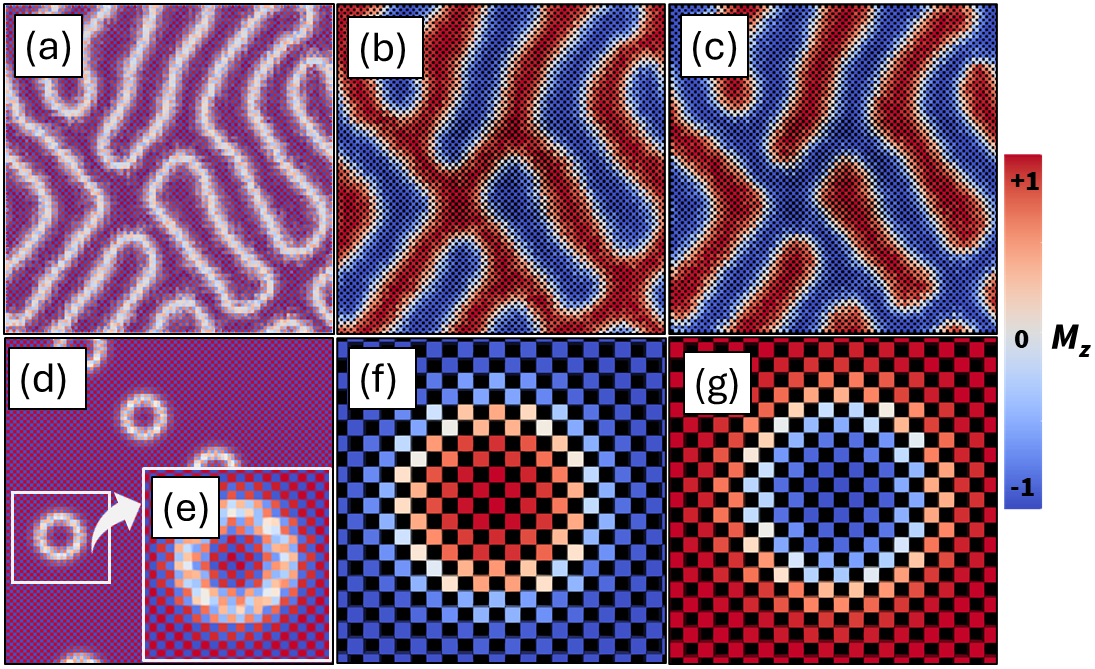}
    \caption{\rj{Magnetization structures formed in the AFM layer. (a) Labyrinth domain state formed at low anisotropy. This can be split into two sublattices, as shown in (b) and (c). (d) Skyrmions formed at a G-type AFM interface. (e) Zoomed view of one of the skyrmionsSkyrmion formed in the (f) sublattice 1 and (g) sublattice 2.}} 
    \label{gtype-afm-skyrmion}
\end{figure}

Skyrmion formations within the G-type AFM structure are depicted in Fig.~\ref{gtype-afm-skyrmion}. This AFM lattice is divisible into two \rj{opposite} sublattices, each being individually ferromagnetic. The antiferromagnetic coupling between the nearest neighbors ensures that these sublattices are essentially mirror images of each other. \rj{Here, the uniaxial anisotropy provides the out-of-plane field needed to stabilize skyrmion structures because using an external field is not feasible. This is because the sublattices are arranged oppositely, meaning an external field in any direction would energetically benefit one sublattice while disadvantaging the other. The magnetization in the layer evolves with the anisotropy strength analogous to the response found in the FM layer to the external field. At low anisotropy, the magnetization is in an AFM-labyrinth state as shown in Fig. \ref{gtype-afm-skyrmion} (a).  When this state is divided into two sublattices and analyzed separately, it reveals that it is made of two FM labyrinth domain patterns which we saw before.  With an increase in the anisotropy field strength, these domains are disrupted and bubble-like structures are formed, these are the AFM-skyrmions.} Each bubble is characterized by a ring-shaped area where the magnetization lies in the plane. Inside and outside this ring, the magnetization adheres to the typical G-type pattern. When the magnetization structure of the bubble is divided and analyzed separately for each sublattice, it reveals the presence of two skyrmions, each with an opposite polarity in its respective sublattice. The discovery of these magnetic configurations in G-type antiferromagnetic (AFM) materials is noteworthy. Numerical simulations have demonstrated that these structures can move at high speeds in AFM lattices \cite{Tresmina2022}. Additionally, the formation of these structures involves the coupling of two skyrmions with opposite topological characteristics. This unique coupling suggests the potential for these skyrmions to neutralize each other's Hall effect, which is a significant aspect of their behavior in magnetic materials.

\section{Conclusion}
In our study, we employed atomistic spin dynamics and Monte Carlo simulations to investigate how various material properties influence the exchange bias field at ferromagnetic FM/AFM interfaces. We observed that the exchange bias field is directly proportional to the interlayer coupling strength, underscoring its critical role. However, it is important to note that our model assumes a perfectly uncompensated interface and does not take into account factors like defects, interlayer diffusion, or inter-layer roughness. These elements could potentially modify the mechanism of exchange bias and may reduce the strength of the bias field, as suggested in previous studies \cite{Stamps2000, Manna_2016, Usadel2010, Dantas2005, Moritz2016}. Future research that incorporates these effects could yield more comprehensive results.
Furthermore, we conducted a detailed analysis of the stability of skyrmions in FM/AFM multilayers. Our findings indicate that skyrmions formed in the FM monolayer via an exchange bias interface exhibit greater stability compared to those stabilized by external fields, as determined through GNEB calculations. \rjb{This is, due to the formation of magnetic imprints of the skyrmion in the AFM layer at the interface.} However, when the number of FM layers was increased to three or more, the exchange bias from the interface was insufficient to maintain the skyrmion stability in the upper layers. This observation aligns with our previous findings, demonstrating an inverse relationship between the exchange bias field and the number of FM layers, and is consistent with the experimental work on exchange bias stabilized skyrmions by Rana et al. \cite{Rana2020}. In their study, skyrmions are stabilized by exchange bias in a ferromagnetic NiFe layer of approximately 0.3 nm thickness, equivalent to two or fewer atomic layers. Future research efforts aiming to extend the exchange bias strength across more layers, thereby stabilizing skyrmion tubes with additional FM layers, could be particularly intriguing.

Additionally, our studies on the current-induced dynamics of isolated skyrmions on monolayers and AFM-coupled skyrmions on bilayers have provided insights into the fundamental mechanisms of skyrmion Hall effects. Based on these insights, we proposed the design of an exchange-stabilized skyrmion conduit. This setup is envisioned to host stable skyrmions in the absence of external fields and facilitate their high-velocity movement, reaching speeds up to 3 \text{Km/s}: \rjb{ which is several times faster than generally reported velocities \cite{Lai2017,Kolesnikov2018vel,Gobel2019,juge2019,pham2024}}. The practical realization of such a setup could mark a significant milestone in the field of spintronics. Nevertheless, further investigations are required, particularly to identify materials with optimal properties for the device's construction. Given our existing knowledge of material parameters \rjb{like $J$, $D$}, etc., density functional theory calculations could be instrumental in identifying suitable materials or material combinations for this purpose.

\section{Acknowledgment}
Financial support from Carl Tryggers stiftelse Grant No. CTS:21-1400, the
Swedish Research Council (Vetenskapsr{\aa}det, VR) Grant No. 2016-05980, Grant No. 2019-05304, and Grant No. 2023-04239, and the Knut and Alice Wallenberg foundation Grant No. 2018.0060, Grant No. 2021.0246, and Grant No. 2022.0108 is acknowledged.  
The China Scholarship Council (CSC) is acknowledged by Z.L. 
The Wallenberg Initiative Materials Science for Sustainability (WISE) funded by the Knut and Alice Wallenberg Foundation is also acknowledged.
The computations/data handling were enabled by resources provided by the National Academic Infrastructure for Supercomputing in Sweden (NAISS), partially funded by the Swedish Research Council through grant agreement no. 2022-06725.

\nocite{ASD2VTK}
\newpage
\newpage

\bibliography{apssamp}
\end{document}